\documentclass[10pt,twocolumn]{article}

\usepackage[a4paper,margin=1.7cm,columnsep=0.7cm]{geometry}
\usepackage[style]{abstract}
\renewcommand{\abstitlestyle}[1]{}
\usepackage{graphicx}
\usepackage{amsmath,amssymb}
\usepackage{float,placeins}
\usepackage{xcolor,booktabs}
\AddToHook{env/figure/begin}{\setlength{\abovecaptionskip}{5pt}}
\AddToHook{env/figure*/begin}{\setlength{\abovecaptionskip}{5pt}}
\usepackage[english]{babel}
\usepackage[numbers,sort&compress]{natbib}
\usepackage{hyperref}
\hypersetup{
	pdftitle={Spin distribution of fission fragments involving transverse vibrational modes},
	pdfauthor={D. E. Lyubashevsky, A. A. Pisklyukov, Yu. D. Shcherbina, T. Yu. Shashkina, P. V. Kostryukov},
	colorlinks=true,
	linkcolor=blue,
	filecolor=magenta,      
	urlcolor=blue,
	citecolor=blue,
	pdfpagemode=UseNone,
}

\makeatletter
\renewcommand\section{\@startsection{section}{1}{\z@}%
  {-3.5ex \@plus -1ex \@minus -.2ex}%
  {2.3ex \@plus .2ex}%
  {\normalfont\large\bfseries\raggedright}}
\makeatother

\makeatletter
\renewcommand{\@maketitle}{%
  \begin{center}
    {\large\bfseries\MakeUppercase{\@title}\par}
    \vskip 1em
    {\normalsize\@author\par}
  \end{center}%
}
\makeatother

\begin{document}
\title{Spin distribution of fission fragments involving transverse vibrational modes}
\author{D.\,E. Lyubashevsky\textsuperscript{1,2}\thanks{\href{mailto:lyubashevskiy@aes.vsu.ru}{lyubashevskiy@aes.vsu.ru}},
A.\,A. Pisklyukov\textsuperscript{1},
Yu.\,D. Shcherbina\textsuperscript{1},
T.\,Yu. Shashkina\textsuperscript{1}, and
P.\,V. Kostryukov\textsuperscript{1}\\[.25ex]
{\small\itshape\textsuperscript{1}Voronezh State University, 1 Universitetskaya Square, Voronezh 394018, Russia}\\[-.2ex]
{\small\itshape\textsuperscript{2}International Institute of Computer Technologies, 29B Solnechnaya Street, Voronezh 394026, Russia}}
\date{}

\twocolumn[{%
  \begin{@twocolumnfalse}
    \maketitle
    \vspace{-0.5em}
    \begin{abstract}
We present a closed-form analytical description of fragment spin distributions in low-energy neutron-induced and spontaneous fission. We assume that the fragment spins and relative orbital angular momentum arise from zero-point bending and wriggling oscillations of the prefragments near the scission point. The model yields the joint spin distribution, the spin distribution for each fragment, and its mean value. Fragment moments of inertia are estimated with a hydrodynamic model using nonequilibrium scission deformations inferred from prompt-neutron multiplicities. For $^{232}\mathrm{Th}(n,f)$, $^{238}\mathrm{U}(n,f)$, and spontaneous fission of $^{252}\mathrm{Cf}$, the calculated mean spins reproduce the observed scale and sawtooth dependence on fragment mass. The zero-point approximation concerns the population of the two transverse modes and is compatible with dissipation during the descent from saddle to scission.
    \end{abstract}
    \vspace{1em}
  \end{@twocolumnfalse}
}]
\saythanks

\newlength{\regulartextfloatsep}
\setlength{\regulartextfloatsep}{\textfloatsep}
\setlength{\textfloatsep}{8pt}
\raggedbottom
\makeatletter
\AddToHookNext{shipout/after}{%
  \global\textfloatsep=\regulartextfloatsep
  \global\let\@textbottom\relax
  \global\let\@texttop\relax
}
\makeatother
\section{INTRODUCTION}\label{sec: intro}

The spin distributions of primary fission fragments (PFFs) set the initial conditions for neutron and $\gamma$-ray emission~\cite{skarsvag1963,gavron1976,wilhelmy1972,morreto1989}, thereby affecting particle multiplicities and angular distributions. They also carry information about the dynamics near the scission point. Fragment spins inferred from experiment are typically $5$--$7\hbar$, well above the spin of the fissioning nucleus at low energy. Their predominantly transverse orientation points to collective motion near scission as a possible source of angular momentum. The mechanism of spin generation remains under discussion.

Calculations of the post-scission Coulomb interaction~\cite{rasmussen1969,randrup2023coulomb} yield spin changes of up to about $2\hbar$, smaller than the observed scale. Theoretical descriptions~\cite{gavron1976,rasmussen1969,shneidman2002,kadmensky2015,bunakov2016} therefore consider collective oscillations near the scission point. The direction of the associated spin components relative to the fission axis distinguishes the longitudinal tilting and twisting modes from the doubly degenerate transverse bending and wriggling modes~\cite{nix1965,dossing1985}. Transport estimates~\cite{randrup2022} give long relaxation times for twisting and tilting, a short time for wriggling, and an intermediate, mass-asymmetry-dependent time for bending. This hierarchy motivates the treatment of the transverse modes in the present low-energy model. Their population is discussed in Sec.~\ref{sbs:transverse-oscillations}.

\begin{figure}[!b]
		\centering
		\includegraphics[width=\linewidth]{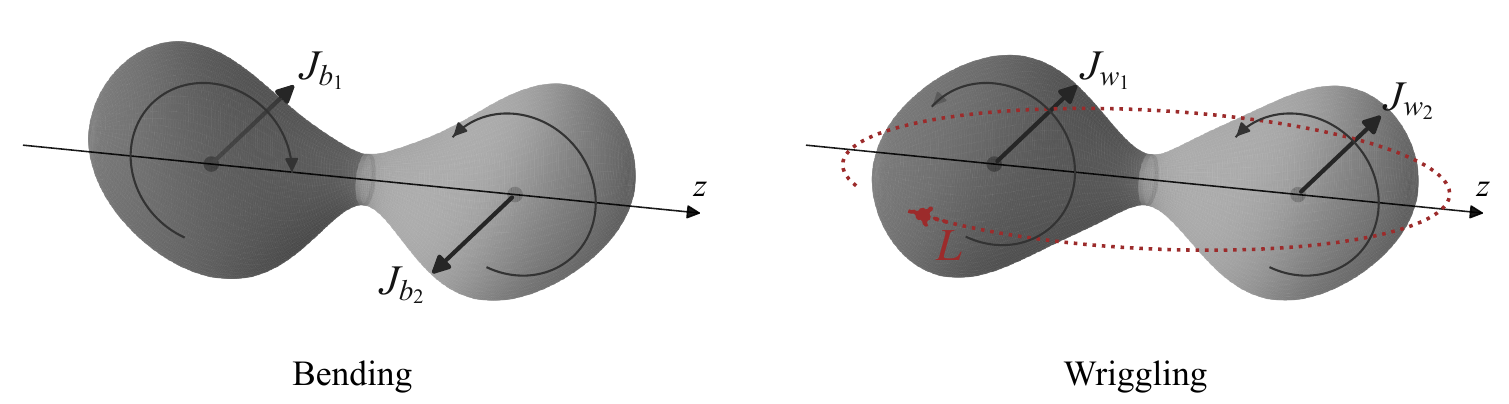}
\caption{Transverse bending and wriggling oscillations of the prefragments. Curved arrows indicate the directions of rotation and straight arrows show the corresponding spin contributions. The fission axis is denoted by $z$.}\label{fig:modes}
	\end{figure}

As shown in Fig.~\ref{fig:modes}, bending corresponds to counter-rotation of the prefragments and gives oppositely directed spin contributions. In wriggling the prefragments rotate in the same direction. Their total spin $J_{\rm w}=J_{1_{\rm w}}+J_{2_{\rm w}}$ is balanced by the relative orbital angular momentum $\mathcal{L}_{\rm w}=-J_{\rm w}$. Both modes conserve the total angular momentum. We evaluate the stiffnesses at a fixed shape near the scission point, with the elongation, mass asymmetry, and neck parameters held fixed. Early work emphasised bending~\cite{rasmussen1969,wilhelmy1972,shneidman2002}. Later calculations~\cite{vogt2021,randrup&vogt2021,randrup2022} indicate that the two transverse modes can make comparable contributions.

The fragment spin data~\cite{wilson2021} for $^{232}\mathrm{Th}(n,f)$, $^{238}\mathrm{U}(n,f)$, and $^{252}\mathrm{Cf}(\mathrm{sf})$ allow several descriptions of spin generation to be compared. Event-by-event statistical models such as \texttt{FREYA}~\cite{verbeke2015,vogt2021,randrup&vogt2021,randrup2022} sample the bending and wriggling angular momenta at finite temperature and follow the subsequent emission of neutrons and $\gamma$ rays. In density functional theory, the spins of primary fragments are generated within the microscopic fission dynamics. Such calculations~\cite{stetcu2022,bulgac2022,li2026intrinsic,marevic2026spins,scamps2026spinparity} yield distributions of spin magnitudes. With all three spatial spin components, one can also study the orientation of the spins relative to the fission axis and correlations between partners~\cite{scamps2023,scamps2024}. Both approaches reproduce important features of the data and rely on different assumptions about excitation and dynamics near scission.

The present study builds on our earlier work~\cite{lyubashevsky2025,lyubashevsky2026,lyubashevsky2025prc} on fission-fragment spins. We obtain the joint and the individual spin distributions of the primary fragments analytically from the momentum-space wave functions of the zero-point bending and wriggling modes. Moments of inertia are calculated from scission deformations inferred from prompt-neutron multiplicities. The calculated distributions and the mass dependence of the mean spins are compared with estimates inferred from measured $\gamma$ cascades~\cite{wilson2021}.

\section{SPIN DISTRIBUTIONS OF PRIMARY FISSION FRAGMENTS}\label{sec:model}

\subsection{Stages of low-energy induced and spontaneous fission}

In low-energy neutron-induced fission, neutron capture forms a compound nucleus at an excitation energy $E^*_{\rm CN}=|B_n|+E_n$, where $B_n$ denotes the neutron binding energy and $E_n$ is the incident-neutron energy. The small recoil correction is neglected. On a nuclear timescale the compound nucleus reaches a resonance containing quasiparticle excitations and collective deformation motion. For an axially symmetric nucleus, the resonance~\cite{bohr_mottelson,kadmensky2005} is characterised by the total angular momentum $J$, its laboratory projection $M$, and its projection $K$ on the symmetry axis. In the Wigner representation~\cite{wigner1955}, the corresponding wave function is written schematically as
	\begin{equation}\label{eq:wignerstate}
		\Psi_K^{JM}=b_0\psi_{0K}^{JM}(\boldsymbol{\beta})+
		\sum_{i\ne0}b_i\psi_{iK}^{JM}(\boldsymbol{\beta}),
	\end{equation}
where $\boldsymbol{\beta}$ denotes the deformation parameters and $\psi_{0K}^{JM}(\boldsymbol{\beta})$ describes the collective deformation state. The remaining terms describe quasiparticle excitations. The transition states at the fission barriers determine which values of $J$, $K$, and parity contribute to the fission channel~\cite{bohr_mottelson}.

\begin{figure}[ht]
		\centering
		\includegraphics[width=\linewidth,trim=0 10bp 0 0,clip]{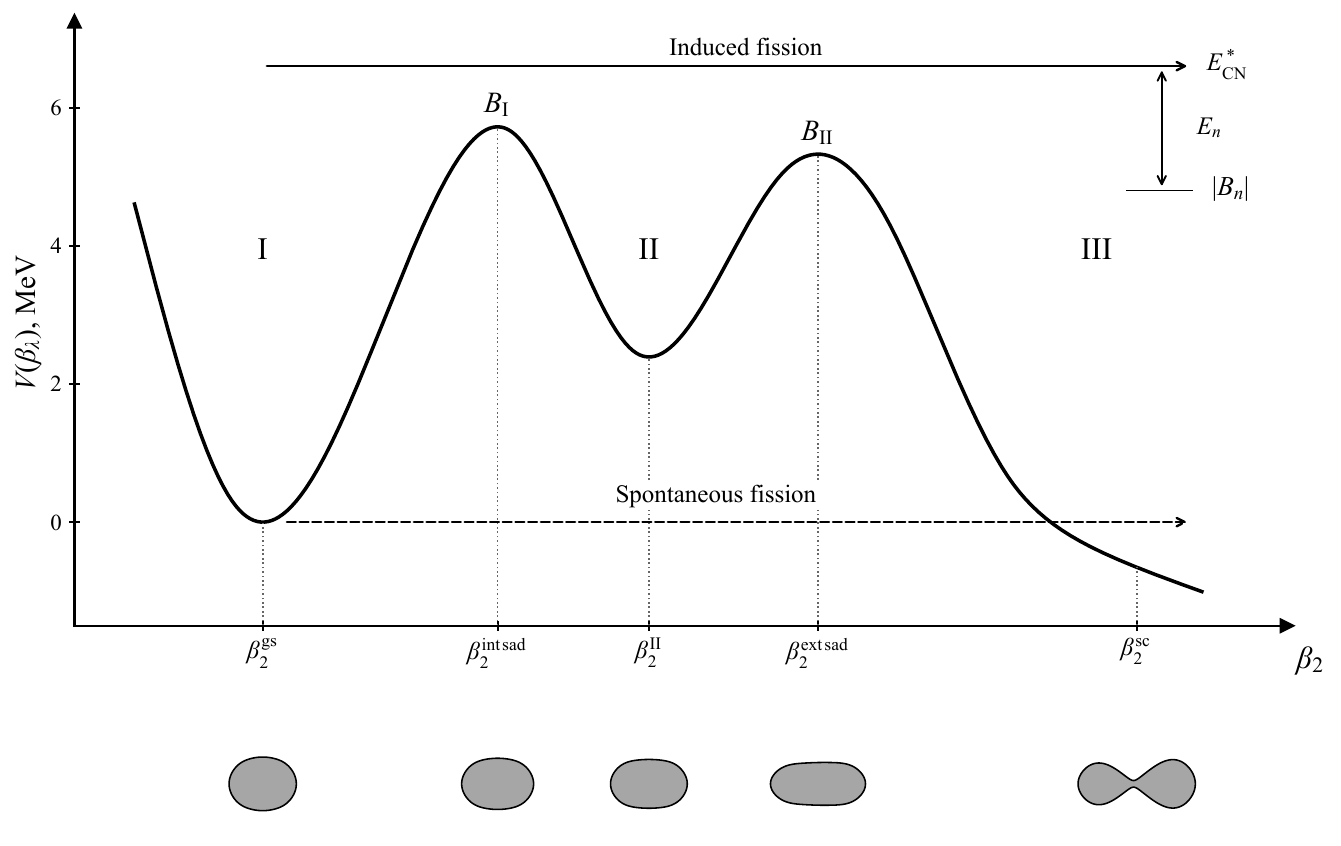}
\caption{Schematic double-humped deformation potential and the evolution of the nuclear shape from the ground state to scission. The solid arrow indicates induced fission at $E^*_{\rm CN}=|B_n|+E_n$. The dashed arrow indicates spontaneous fission by tunnelling through the barriers.}\label{fig:barrier}
	\end{figure}

The subsequent evolution of the selected collective state is illustrated by the double-humped deformation potential~\cite{strutinsky1967} in Fig.~\ref{fig:barrier}. Region I contains the ground-state well and region II the second, isomeric well. The outer barrier separates the second well from region III, which leads to the scission point. In induced fission, the energy supplied by neutron capture allows the collective component of the compound-nucleus state to pass through or over the barriers.

In the picture of a predominantly collective descent~\cite{kadmensky2002,kadmensky2003,kadmensky2005,kadmensky2008}, most of the available energy beyond the outer barrier remains in deformation degrees of freedom and is stored in the nonequilibrium shape of the fissioning nucleus. In this sense a cold fissioning system can carry substantial collective deformation energy while its intrinsic thermal excitation remains small.

A related test comes from angular asymmetries of different $P$ and $T$ parities in low-energy fission. In a quantum-mechanical description of these asymmetries~\cite{kadmensky2005,kadmensky2014asym,kadmensky2016asym}, $K$ is retained as a characteristic of the transition state during the descent from the outer barrier. Substantial intrinsic heating would increase Coriolis mixing of the $K$ components and weaken the dependence of the asymmetries on the initial state. The observed patterns are consistent with a cold descent.

Spontaneous fission of an even--even nucleus starts from the collective ground state with $J=M=K=0$. The system reaches region III by tunnelling through the barriers. In the cold-fission picture~\cite{kadmensky2005}, most of the energy released between the outer barrier and the scission point is stored in the nonequilibrium deformations of the prefragments. In this limit the two deformed prefragments reach scission with little intrinsic thermal excitation. We consider their bending and wriggling oscillations below.

Calculations of the collective angular motion of touching fragments~\cite{shneidman2025angular,rahmatinejad2025tke} show that deformation energy can dominate for selected scission configurations in the spontaneous fission of $^{252}\mathrm{Cf}$. The partition between nonequilibrium deformation energy and intrinsic thermal excitation depends on the fragment pair and the assumed configuration. A strongly dissipative calculation~\cite{albertsson2020} also finds configurations in thermal-neutron-induced fission for which intrinsic excitation exceeds the deformation energy released. We therefore treat deformation dominance as a model assumption and test the sensitivity of the predicted spins to finite temperature.

\subsection{Transverse bending and wriggling modes}\label{sbs:transverse-oscillations}
	
We describe the spin distributions near the scission point in terms of transverse bending and wriggling. After separating the overall rotation, the harmonic Hamiltonian~\cite{nix1965,kadmensky2017} for each mode $t={\rm w},b$ is
	\begin{equation}\label{eq:modehamiltonian}
		\begin{aligned}
			\hat H_t&=\sum_{p=x,y}\left[
			\frac{\hat J_{t_p}^{\,2}}{2I_t}
			+\frac{\kappa_t q_{t_p}^{\,2}}{2}\right],\\
			\hat J_{t_p}&=-i\hbar\frac{\partial}{\partial q_{t_p}},
		\end{aligned}
	\end{equation}
where $q_{t_p}$ are the dimensionless internal normal coordinates of the two equivalent transverse components. At finite effective temperature $T$, the oscillator eigenstates are thermally populated. The corresponding momentum-space probability density is
	\begin{equation}\label{eq: 1}
		P\!\left(J_{t_x},J_{t_y}\right)
		=\frac{1}{\pi C_t(T)}
		\exp\!\left[-\frac{J_{t_x}^2+J_{t_y}^2}{C_t(T)}\right].
	\end{equation}
The temperature dependence of the width is
	\begin{equation}\label{eq:CfiniteT}
		C_t(T)=I_t\hbar\omega_{0_t}\coth\!\left(\frac{\hbar\omega_{0_t}}{2T}\right).
	\end{equation}
Here $I_t$ is the moment of inertia and $\omega_{0_t}$ is the frequency of the mode. In the limit $T\ll\hbar\omega_{0_t}$, $C_t(T)\simeq I_t\hbar\omega_{0_t}$. For $T\gg\hbar\omega_{0_t}$, $C_t(T)\simeq 2I_tT$.
For either transverse component in the ground state of $\hat H_t$,
	\begin{equation}\label{eq:zeropointvariance}
		\langle\hat J_{t_p}\rangle_0=0,\qquad
		\langle\hat J_{t_p}^{\,2}\rangle_0
		=\frac{I_t\hbar\omega_{0_t}}{2},
	\end{equation}
so the sum of the two transverse variances is $C_t(0)=I_t\hbar\omega_{0_t}$. The baseline calculation assumes that the additional excitation of each mode is small compared with its quantum $\hbar\omega_{0_t}$ and uses the oscillator ground states at the adopted scission configuration.

The mode frequency is given by
	\begin{equation}\label{eq:modefrequency}
		\omega_{0_t}=\sqrt{\frac{\kappa_t}{I_t}}.
	\end{equation}
The stiffness $\kappa_t$ is defined with respect to either of the two equivalent dimensionless normal coordinates $q_t$, with a kinetic energy of $I_t\dot{q}_t^{2}/2$. At a fixed scission configuration, the stiffness entering Eq.~\eqref{eq:modefrequency} is the local curvature of the total energy $E_t(q_t)$ at equilibrium,
	\begin{equation}\label{eq:modestiffness}
		\kappa_t=\left.\frac{d^2E_t}{dq_t^2}\right|_{q_t=0}.
	\end{equation}
In Eq.~\eqref{eq:modestiffness}, the normal coordinate of the selected mode is varied~\cite{nix1965} at fixed elongation, mass asymmetry, and neck parameters. The resulting curvature characterises the connected configuration. The Coulomb interaction between separated fragments~\cite{randrup2023coulomb} acts at a later stage.
	
The bending frequencies are taken from calculations~\cite{shneidman2002} for eight fragment pairs in spontaneous fission of $^{252}\mathrm{Cf}$. The experimental sample~\cite{wilson2021} covers 31 pairs. The calculated pairs span light-fragment masses from $100$ to $120$ and heavy-fragment masses from $132$ to $152$, covering most of the range studied here. For the other pairs we interpolate or extrapolate the frequencies using their approximately linear dependence on fragment mass $A_f$ over these intervals.

For neutron-induced fission of $^{232}\mathrm{Th}$ and $^{238}\mathrm{U}$, the bending frequencies are obtained from the $^{252}\mathrm{Cf}(\mathrm{sf})$ mass interpolation over the corresponding fragment mass ranges. The transfer introduces a systematic uncertainty because the stiffness, compound-nucleus mass, and fission trajectory vary between these systems.

Following the frequency ratios of Ref.~\cite{nix1965}, we use $\omega_{0_{\rm w}}/\omega_{0_b}=2.6$ for $^{252}\mathrm{Cf}(\mathrm{sf})$ and $2.5$ for neutron-induced fission of $^{232}\mathrm{Th}$ and $^{238}\mathrm{U}$. The bending frequencies and the frequency ratios are model inputs fixed independently of the spin data. The values used are given in Tables~\ref{tab:t1}--\ref{tab:t3}. Direct calculations of $\kappa_t$ for different mass splits and fissioning systems are left for future work.
	
\begin{figure}[htbp]
		\centering
		\includegraphics[width=0.98\linewidth,trim=0 6bp 0 0,clip]{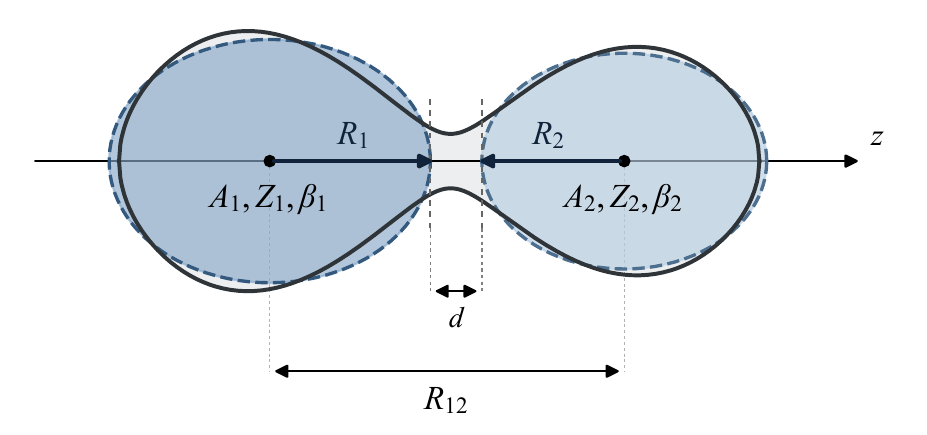}
\caption{Geometry at the scission point. Dashed lines represent the spheroidal surfaces of the prefragments $(A_1,Z_1,\beta_1)$ and $(A_2,Z_2,\beta_2)$. The solid line traces the profile of the fissioning nucleus.}\label{fig:neck}
		\vspace{-1em}
	\end{figure}

In the adopted description of the wriggling mode~\cite{randrup&vogt2021,vogt2021,randrup2022}, the moment of inertia $I_{\rm w}$ is
	\begin{equation}\label{eq:2}
		I_{\rm w}=\frac{I_0}{I} (I_1+ I_2), 
	\end{equation}
where $I_i$ is the inertia of prefragment $i$, $I_0=\mu R_{12}^2$ is the orbital inertia, and $\mu=M_1M_2/(M_1+M_2)$ is the reduced mass. Here $M_i$ is the mass of prefragment $i$, and $I=I_0+I_1+I_2$. The centre-to-centre distance along the fission axis is $R_{12}=R_1+R_2+d$, where $R_1$ and $R_2$ are the longitudinal radii and $d$ is the surface separation (the neck length in Fig.~\ref{fig:neck}). We adopt an effective gap of $4$~fm for all mass splits, following the scission geometry of Ref.~\cite{randrup2022}. Establishing its consistency with the deformations and transferred frequencies would require a joint calculation of the geometry, stiffnesses, and frequencies.
	
The longitudinal radius $R_i$ for a quadrupole deformation $\beta_{2_i}$ is given by
\begin{equation}\label{eq:fragmentradius}
 R_i(\beta_{2_i})=r_0 A_i^{1/3}
 \left[1-\frac{\beta_{2_i}^2}{4\pi}
       +\beta_{2_i}\sqrt{\frac{5}{4\pi}}\right].
\end{equation}
An alternative bending inertia used in the same studies is defined by the reduced moment of inertia of the two prefragments,
	\begin{equation}\label{eq:3} 
		I'_b=\frac{I_1 I_2}{I_1 + I_2}.
	\end{equation}
Equation~\eqref{eq:3} describes the relative rotation of the two prefragments. In the numerical calculations below we follow the definition of Ref.~\cite{shneidman2015}, in which the heavy prefragment rotates relative to the remainder of the fissioning nucleus. The corresponding reduced inertia is
	\begin{equation}\label{eq:4}
		I_b =\frac{I_0 I_H}{I_0 + I_H},
	\end{equation}
where $I_H$ is the inertia of the heavy fragment and $I_0$ is the orbital inertia of the complementary part. This $I_b$ must be used with the corresponding stiffness and frequency. Both $I'_b$ in Eq.~\eqref{eq:3} and $I_b$ in Eq.~\eqref{eq:4} assume two developed prefragments. We restrict the analysis to the experimental mass range~\cite{wilson2021}.
	
The deformations $\beta_{2_i}$ entering these inertias can substantially exceed their ground-state equilibrium values for near-magic fragments, including Sn isotopes. In the scission-point model of Ref.~\cite{wilkins1976}, which includes shell effects at finite deformation, equilibrium shapes near mass $130$ are nearly spherical owing to the $Z=50$ and $N=82$ shell closures. The $\beta_{2_i}$ used here describe nonequilibrium prefragment shapes at the scission point. Some of the stored deformation energy is released after scission and contributes to neutron and $\gamma$-ray emission. We estimate the deformation energy from the total excitation energy $TXE$ available for fragment de-excitation.

The systematics of Grudzevich~\cite{grudzevich2000} relate $TXE$ to the prompt-neutron multiplicity $\nu$ by
	\begin{equation}\label{eq:Unu}
		TXE = 5 + 4\nu + \nu^2.
	\end{equation}
This empirical relation and the parametrisation $U=7\left(\nu+\tfrac{3}{7}\right)$ of D{\o}ssing et al.~\cite{dossing2024} give a first estimate of the excitation energy used to infer the nonequilibrium deformations. For $^{238}\mathrm{U}(n,f)$ we use the Point-by-Point multiplicities~\cite{tudora2016} near $E_n=2$~MeV. The estimate for $^{232}\mathrm{Th}(n,f)$ follows Ref.~\cite{tudora2013} at a similar energy. For $^{252}\mathrm{Cf}(\mathrm{sf})$ we use the mass-dependent measurements of Walsh and Boldeman~\cite{walsh1977}, as in Ref.~\cite{lyubashevsky2025}. The model estimates cover masses for which direct measurements are incomplete. For an individual fragment the energy balance is approximated by
	\begin{equation}\label{eq:energybalance}
		TXE\simeq E_{\rm int}^{(\rm sc)}
		+E_{\rm def}^{(\rm rel)},
	\end{equation}
where $E_{\rm int}^{(\rm sc)}$ is the intrinsic excitation at scission and $E_{\rm def}^{(\rm rel)}$ is the energy released as the fragment relaxes to its equilibrium shape. To obtain a first estimate of the deformation, we set $E_{\rm int}^{(\rm sc)}$ to zero and attribute $TXE$ to deformation relaxation. A finite intrinsic component would reduce the energy available for deformation and could lower the inferred $\beta_2$ and moment of inertia. This uncertainty enters the reconstruction of the shape, whereas the thermal check below concerns the population of the transverse modes at a given shape. Reference~\cite{albertsson2020} illustrates a different energy partition. In the liquid-drop model~\cite{strutinsky1967} the deformation energy relative to the spherical shape is
	\begin{equation}\label{eq:Udef}
		TXE_{\rm def} = \sigma A^{2/3}\left[\frac{2}{5}(1-x)\eta^2 - \frac{4}{105}(1+2x)\eta^3\right],
	\end{equation}
where $\sigma=16$~MeV, $x=Z^2/(45A)$ is the fissility parameter, and $\eta=\tfrac23\beta_2$. We subtract the equilibrium deformation energy obtained from the ground-state shapes~\cite{moller2016} and solve Eq.~\eqref{eq:Udef} for the nonequilibrium $\beta_2$ of each fragment. The neutron multiplicities $\nu(A)$ measured by Walsh and Boldeman~\cite{walsh1977} imply appreciable nonequilibrium deformations even near mass $130$, where the equilibrium shapes of near-magic fragments are nearly spherical. For example, the measured multiplicity $\nu\simeq0.36$ for $^{130}\mathrm{Sn}$ constrains its deformation energy. The multiplicities fix the deformation inputs independently of the spin data. The largest deformations, $\beta_2\gtrsim0.9$ for a few pairs, are near the limit of applicability of the low-order expansion in Eq.~\eqref{eq:Udef} and should be treated as estimates.

At a given quadrupole deformation, the moment of inertia is estimated in the hydrodynamic model~\cite{bohr_mottelson,sitenko2014}. For irrotational flow, it is
	\begin{equation}\label{eq:Ihydro}
		I = \frac{9 M R^2}{4\pi}\,
		\frac{\beta_2^2\left(1 + \tfrac{1}{4}\sqrt{\tfrac{5}{4\pi}}\,\beta_2\right)^2}
		{2 + \sqrt{\tfrac{5}{4\pi}}\,\beta_2 + \tfrac{25}{16\pi}\beta_2^2},
	\end{equation}
which to leading order in $\beta_2$ gives the ratio to the rigid-body inertia
	\begin{equation}\label{eq:Iratio}
		\frac{I}{I_{\rm rigid}} \simeq \frac{45}{16\pi}\,\beta_2^2 .
	\end{equation}
The hydrodynamic model gives appreciably reduced moments of inertia. The tabulated values follow Refs.~\cite{lyubashevsky2025,lyubashevsky2026}, and Eqs.~\eqref{eq:Ihydro} and \eqref{eq:Iratio} illustrate the scale of the reduction. Near-magic fragments have $I/I_{\rm rigid}\approx0.1$--$0.3$, while more strongly deformed fragments away from closed shells reach $I/I_{\rm rigid}\approx0.5$--$0.7$. The \texttt{FREYA} parametrisation~\cite{randrup2022} also reduces the inertia near magic numbers. Here the inertia depends on the inferred nonequilibrium deformation, so the same isotope can have different scission inertias when produced by different fissioning nuclei. Shell corrections beyond the liquid-drop estimate of Eq.~\eqref{eq:Udef}, along the lines of Ref.~\cite{wilkins1976}, could refine these results.
	
The transverse fragment spin projections are expressed in terms of the mode angular momenta $J_{t_p}$, with $p=x,y$. In the transformation of Ref.~\cite{randrup&vogt2021}, each fragment's share of the wriggling angular momentum is weighted by its moment of inertia,
	\begin{equation}\label{eq:5}
		J_{i_{p}}=\frac{I_i}{I_1 + I_2}J_{\text{w}_{p}}+(-1)^{i+1}J_{b_{p}}.	
	\end{equation}
Equation~\eqref{eq:5} gives the wriggling projection $J_{{\rm w}_p}$ as the sum of the two spin projections,
	\begin{equation}\label{eq:6}
		J_{\text{w}_{p}}= J_{1_{p}}+ J_{2_{p}}.
	\end{equation}
The corresponding bending projection $J_{b_p}$ follows from Eqs.~\eqref{eq:5} and \eqref{eq:6},
	\begin{equation}\label{eq:7}
		J_{b_{p}} = J_{1_{p}} -\frac{I_1}{I_1 +I_2}J_{\text{w}_{p}}=
		\frac{I_2 J_{1_{p}} - I_1 J_{2_{p}}}{I_1+I_2}.
	\end{equation}
For statistically independent bending and wriggling oscillations the joint distribution of the mode angular momenta~\cite{kadmensky2017} factorises,
	\begin{equation}\label{eq:8}
		P(J_{ \text{w}_{p}}, J_{b_{p}}) = P(J_{\text{w}_{p}}) P(J_{b_{p}}).
	\end{equation}
The transformation in Eqs.~\eqref{eq:5}--\eqref{eq:7} gives the joint distribution of the primary fragment spin projections,
\begin{equation}\label{eq:9}
\begin{aligned}
&P(J_{1_{x}}, J_{2_{x}}, J_{1_{y}}, J_{2_{y}}) \\
&\quad=\frac{1}{\pi^2 C_{\rm w} C_b}
   \exp\!\left[-\frac{J_{\mathrm{w}_x}^2+J_{\mathrm{w}_y}^2}{C_{\rm w}}
               -\frac{J_{b_x}^2+J_{b_y}^2}{C_b}\right] \\
&\qquad\times
 \left|\frac{\partial(J_{\mathrm{w}_x}, J_{b_x}, J_{\mathrm{w}_y}, J_{b_y})}
              {\partial(J_{1_x}, J_{2_x}, J_{1_y}, J_{2_y})}\right| \\
&\quad=\frac{1}{\pi^2 C_{\rm w} C_b}\exp\Biggl[
 -\frac{(J_{1_x}+J_{2_x})^2+(J_{1_y}+J_{2_y})^2}{C_{\rm w}} \\
&\qquad
 -\frac{(I_2J_{1_x}-I_1J_{2_x})^2+(I_2J_{1_y}-I_1J_{2_y})^2}
       {C_b(I_1+I_2)^2}
 \Biggr].
\end{aligned}
\end{equation}
We next express Eq.~\eqref{eq:9} in polar coordinates. In the transverse plane the polar angles $\theta_1$ and $\theta_2$ are fixed at $\pi/2$. We replace the azimuthal angles $\varphi_1$ and $\varphi_2$ by the relative angle $\varphi=\varphi_1-\varphi_2$ and the mean angle $\varphi'=(\varphi_1+\varphi_2)/2$. Integrating over $\varphi'$ gives
	\begin{equation}\label{eq:10}
		\begin{aligned}
			&P(J_1, J_2, \varphi) = \frac{2 J_1 J_2}{\pi C_{\rm w} C_b} \exp \Bigl[ -J_1^2(\alpha I_2^2 + C_{\rm w}^{-1}) \\ 
			&- J_2^2(\alpha I_1^2 + C_{\rm w}^{-1})  +2 J_1 J_2 \cos{\varphi}(\alpha I_1 I_2 - C_{\rm w}^{-1} )\Bigr],
		\end{aligned}
	\end{equation}
where $\alpha=C_b^{-1}(I_1+I_2)^{-2}$.
Integrating Eq.~\eqref{eq:10} over the spin magnitude of one fragment and over $\varphi\in[0,2\pi)$ gives the normalised distribution $P(J_i)$ for the spin magnitude of the other fragment,
	\begin{equation}\label{eq:11}
		P(J_i) = \frac{2 J_i}{\delta_i} \exp \left[ - \frac{J_i^2}{\delta_i} \right],
	\end{equation}
where $\delta_i=I_i^2 C_{\rm w} (I_1+I_2)^{-2}+C_b$.

The mean spin is
	\begin{equation} \label{eq:12}
\begin{aligned}
 \langle J_i\rangle &= \int_0^\infty P(J_i)J_i\,dJ_i \\
 &= \int_0^\infty \frac{2J_i^2}{\delta_i}\,
       e^{-J_i^2/\delta_i}\,dJ_i = \frac{\sqrt{\pi\delta_i}}{2}.
\end{aligned}
	\end{equation}
\begin{figure*}[!t]
\centering
\includegraphics[width=\textwidth,trim=0 5.5bp 0 0,clip]{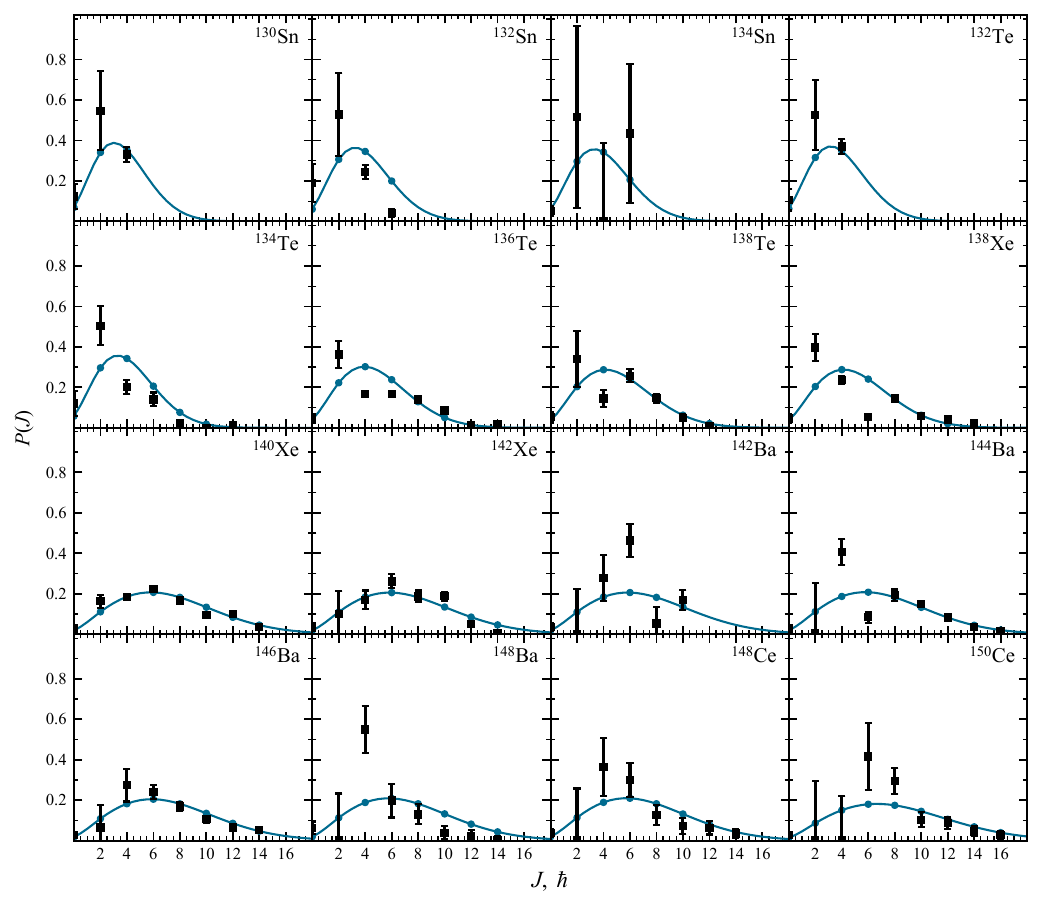}
\caption{Spin distributions of heavy fragments in $^{238}\mathrm{U}(n,f)$. Black squares with error bars show the experimental distributions~\cite{wilson2021} reconstructed from $\gamma$ cascades. Solid lines show the primary-fragment calculation based on Eq.~\eqref{eq:11}.}\label{fig:f4}
\end{figure*}

The two transverse contributions to $\delta_i$ set the scale of the individual spin distribution and its mean. Tables~\ref{tab:t1}--\ref{tab:t3} give the mode inertias and calculated mean spins for $^{238}\mathrm{U}(n,f)$, $^{232}\mathrm{Th}(n,f)$, and $^{252}\mathrm{Cf}(\mathrm{sf})$. We compare the distributions and means from Eqs.~\eqref{eq:11} and \eqref{eq:12} with the available data below.

\section{DISCUSSION}\label{sec:disc}
\raggedbottom

\subsection{Comparison with experiment}

We first compare the shapes of the spin distributions. Wilson et al.~\cite{wilson2021} report discrete spin distributions for heavy fragments in $^{238}\mathrm{U}(n,f)$.

The experimental distributions were inferred from $\gamma$ cascades observed after prompt neutron emission. Figure~\ref{fig:f4} compares them with the calculated primary-fragment distributions. Equation~\eqref{eq:11} sets the form of the calculated density. The width parameter $\delta_i$ follows from Eq.~\eqref{eq:12} using the previously calculated mean spins~\cite{lyubashevsky2025,lyubashevsky2026}. We use these widths directly in the comparison.

The probability for an integer spin is obtained by integrating the density over the two adjoining intervals of width $\hbar/2$ within $J\geq0$. Following the presentation of the experimental data, each odd-spin probability is divided equally between the two adjacent even spins.

\begin{figure*}[!t]
\centering
\includegraphics[width=\textwidth,trim=0 5.5bp 0 0,clip]{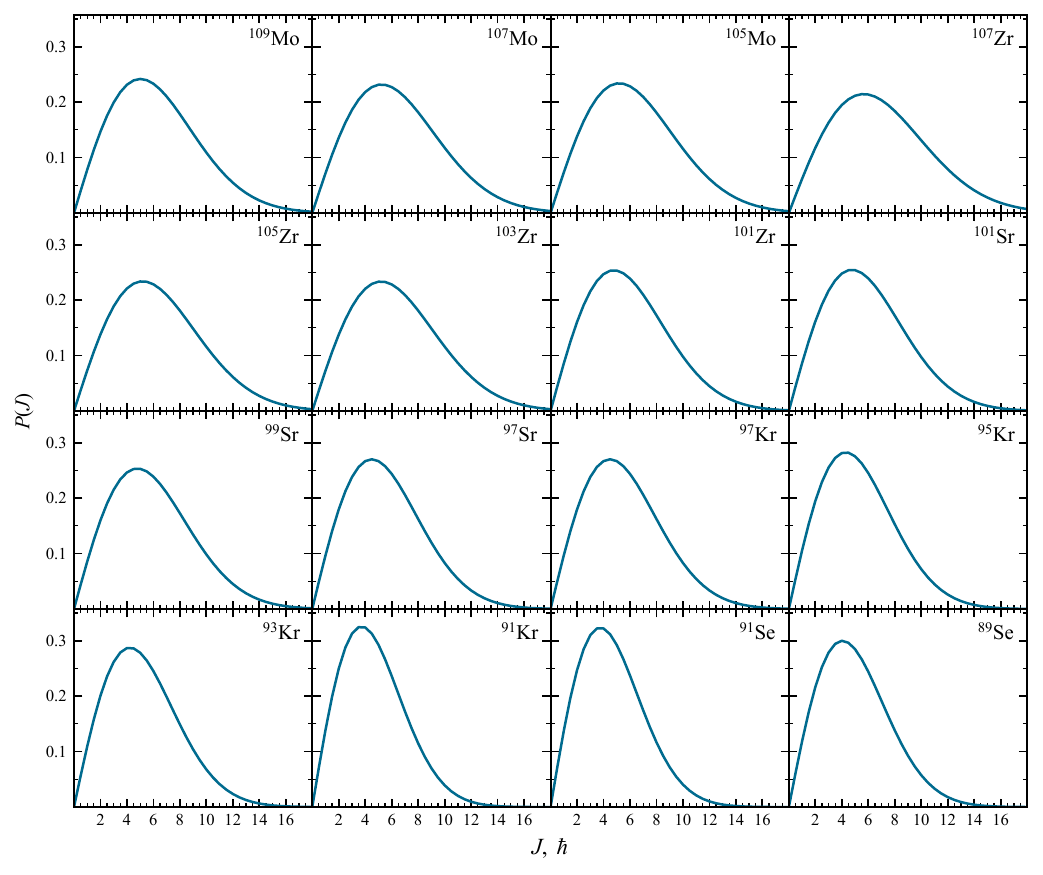}
\caption{Calculated spin distributions of the light primary fragments in $^{238}\mathrm{U}(n,f)$ complementary to the heavy fragments in Fig.~\ref{fig:f4}.}\label{fig:lightdist}
\end{figure*}

The calculation captures the general shape of the spin distributions, with varying agreement across isotopes. At $J=6$, the calculated probability for $^{140}\mathrm{Xe}$ is $0.207$, close to the observed $0.225\pm0.012$. For $^{134}\mathrm{Te}$ at $J=2$, it gives $0.296$ compared with $0.504\pm0.096$. The discrepancies are also appreciable for some Ba isotopes. For $^{148}\mathrm{Ba}$ at $J=4$, the calculation gives $0.188$ compared with $0.549\pm0.117$. The interpretation of the Te and Ba discrepancies should account for spin changes during prompt neutron and statistical $\gamma$-ray emission~\cite{wilson2021,stetcu2021}. Quantifying this effect requires following the fragment de-excitation. Figure~\ref{fig:lightdist} shows the predicted distributions for the light partners.

\begin{figure}[!t]
\centering
\includegraphics[width=\linewidth,trim=4bp 6bp 4bp 4bp,clip]{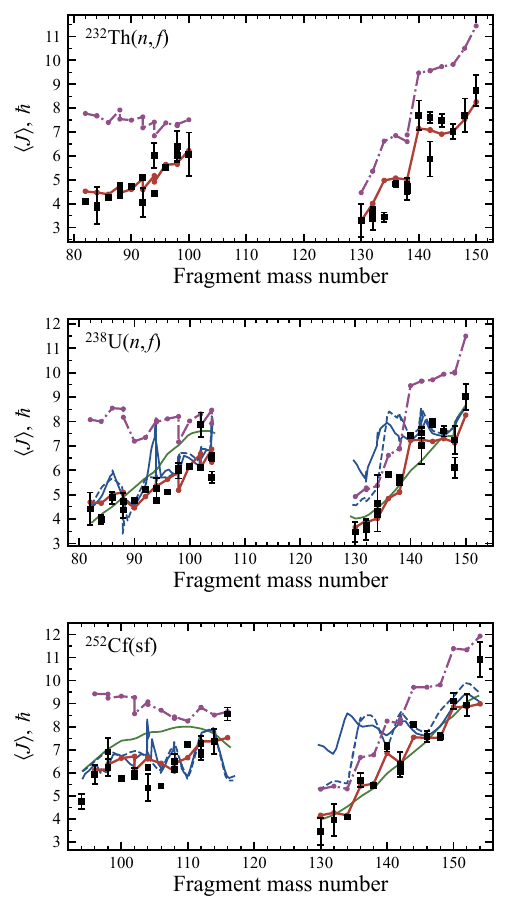}
\caption{Mean angular momenta of the fission fragments. Squares with error bars show the data~\cite{wilson2021}. Red curves show $T=0$ and purple dash-dotted curves $T=1$~MeV. Green curves show FREYA~\cite{randrup&vogt2021} with modified inertias. Blue curves show CGMF~\cite{stetcu2021} after statistical emission, with dashes for the gate on the final $2^+\!\to\!0^+$ transition.}\label{fig:f5}
\end{figure}

Figure~\ref{fig:f5} compares the mean spins in Tables~\ref{tab:t1}--\ref{tab:t3} with the data~\cite{wilson2021} for all three reactions. The zero-point calculation captures the overall scale and the sawtooth mass dependence, including the low values near the Sn isotopes and the subsequent rise along the heavy branch. For the isotopes common to calculation and experiment, the median absolute deviations are $0.42$, $0.30$, and $0.27$ in units of $\hbar$ for $^{232}\mathrm{Th}(n,f)$, $^{238}\mathrm{U}(n,f)$, and $^{252}\mathrm{Cf}(\mathrm{sf})$, respectively. This agreement, obtained with the input parameters fixed independently of the data in Fig.~\ref{fig:f5}, supports the proposed mechanism of spin generation.

In $^{232}\mathrm{Th}(n,f)$, the calculated mean spins are $3.28$ for $^{130}\mathrm{Sn}$ and $7.15$ for $^{140}\mathrm{Xe}$. The measured values are $3.29\pm0.68$ and $7.70\pm0.62$, respectively. For the intervening isotope $^{134}\mathrm{Te}$, the calculated mean of $4.97$ exceeds the observed value of $3.44\pm0.20$.

The comparison with FREYA~\cite{randrup&vogt2021} and CGMF~\cite{stetcu2021} in Fig.~\ref{fig:f5} is particularly revealing near the low-spin end of the heavy branch. At $A_f=130$ the FREYA curves give about $4\hbar$ for both $^{238}\mathrm{U}(n,f)$ and $^{252}\mathrm{Cf}(\mathrm{sf})$, close to our results and to the data. FREYA obtains this minimum through the decrease in the moment of inertia near the Sn isotopes. In our calculation, the corresponding change in inertia enters through the fragment deformations inferred from neutron multiplicities. On the light branch near $A_f=100$, the FREYA curves lie above our results and the data for both reactions.

In CGMF~\cite{stetcu2021}, selecting cascades that end with the $2^+\!\to\!0^+$ transition lowers the calculated means near the Sn isotopes and improves agreement there. Farther along the heavy branch, the deviations are often smaller. For example, near $A_f=140$ in $^{238}\mathrm{U}(n,f)$, both CGMF curves and our result lie near the measured value, while FREYA lies lower. The two CGMF curves include statistical de-excitation and the $1\hbar$ correction used in the original comparison. The sawtooth pattern can therefore reflect both the mass dependence of the moments of inertia at scission and the way in which the subsequent decay selects and changes the spins sampled by the measurement.

\subsection{Contributions of bending and wriggling}

For the factorised distribution in Eq.~\eqref{eq:8}, the mode angular momenta have zero means and contribute additively to the mean-square transverse spin. Equation~\eqref{eq:5} gives
\begin{equation}\label{eq:modalcontributions}
 \langle J_i^2\rangle=C_b+w_i^2C_{\rm w},
 \qquad w_i=\frac{I_i}{I_1+I_2}.
\end{equation}
The last two columns of Tables~\ref{tab:t1}--\ref{tab:t3} give the dominant contribution for each fragment. In wriggling, the two prefragments rotate with the same angular velocity.

Wriggling dominates for the heavy partner in all the mass splits considered with $A_H-A_L\geq50$. In the $^{82}\mathrm{Ge}+{}^{157}\mathrm{Nd}$ split, bending accounts for almost the entire contribution to Ge, while wriggling accounts for about $70\,\%$ for Nd. As the fragment masses approach one another, the scission deformations play a greater role. In $^{109}\mathrm{Mo}+{}^{130}\mathrm{Sn}$, for instance, the near-magic Sn fragment has the smaller inertia and is bending-dominated. In $^{112}\mathrm{Pd}+{}^{140}\mathrm{Te}$, the inertias are similar and so are the modal fractions for the two partners.

Mass asymmetry primarily affects the balance between the two modes for the light fragment. In the strongly asymmetric configurations considered here, the heavier fragment has a larger moment of inertia and carries most of the wriggling angular momentum. Bending makes the same contribution to the mean-square spin of both partners and dominates for the light fragment because its wriggling contribution is small.

The fraction of strongly asymmetric configurations among the selected pairs decreases from thorium to californium. This is consistent with the fragment mass distributions, with asymmetric channels predominant for thorium and nearly symmetric configurations more prominent for californium. Across this sequence the mean wriggling fraction increases for light fragments and changes little for heavy fragments. Giving each fragment equal weight, bending is slightly dominant for thorium, the two contributions are close for uranium, and wriggling is slightly dominant for californium. The average balance thus depends on the reaction, primarily through the changing modal contributions to the light fragments.

\subsection{Zero-point oscillations and intrinsic excitation}

We examine thermal population of the transverse modes for $^{88}\mathrm{Se}+{}^{151}\mathrm{Ce}$ in $^{238}\mathrm{U}(n,f)$ and $^{130}\mathrm{Sn}+{}^{122}\mathrm{Cd}$ in the spontaneous fission of $^{252}\mathrm{Cf}$. We retain the baseline geometry and mode frequencies. The temperature $T$ in Eq.~\eqref{eq:CfiniteT} controls the mode populations at the scission point, with $T=0$ corresponding to the zero-point limit. In thermal equilibrium with the intrinsic degrees of freedom, the intrinsic energy at scission can be estimated from
\begin{equation}\label{eq:temperatureestimate}
 E_{\rm int}^{(\rm sc)}=(a_1+a_2)T^2,
 \qquad a_i=\frac{A_i}{10}.
\end{equation}
In this expression $E_{\rm int}^{(\rm sc)}$ and $T$ are in MeV. For example, $T=1$~MeV corresponds to about $24$--$25$~MeV of intrinsic energy already at scission. The full excitation energy of the separated fragments also includes the energy released as their deformations relax.

Thermal population changes $\delta_i(T)=C_b(T)+w_i^2C_{\rm w}(T)$ while preserving the functional form of Eq.~\eqref{eq:11}. The moments are
\begin{equation}\label{eq:thermalstatistics}
 \langle J_i^n\rangle_T=\delta_i(T)^{n/2}\,
 \Gamma\!\left(1+\frac{n}{2}\right).
\end{equation}
As $\delta_i(T)$ increases, the distribution broadens and its peak moves to higher angular momentum and decreases in height. The $T=1$~MeV curves in Fig.~\ref{fig:f5} were obtained by multiplying the tabulated means by $\sqrt{\delta_i(T)/\delta_i(0)}$, using the mode fractions in Tables~\ref{tab:t1}--\ref{tab:t3}.

At $T\simeq0.2$~MeV, which corresponds to about $1$~MeV of intrinsic excitation, the mean angular momenta of the four fragments in the two examples change by less than $3\,\%$. At $T=1$~MeV, the heavy branch retains its sawtooth shape and shifts to higher angular momenta. For the light branch, the shape of the mass dependence changes as well. Across the three reactions, the median increases range from $36$ to $56\,\%$ for the light fragments and from $28$ to $35\,\%$ for the heavy fragments. The fragments with a large contribution from the lower-frequency bending mode are particularly sensitive to thermal population.

At $T=1$~MeV the predicted angular momenta exceed the measured values along almost the entire light branch. For $^{116}\mathrm{Pd}$ from spontaneous fission of $^{252}\mathrm{Cf}$ and $^{102}\mathrm{Mo}$ from $^{238}\mathrm{U}(n,f)$, the thermal correction brings the calculation closer to the data~\cite{wilson2021}. This improvement could reflect thermal excitation or the specific scission deformations and moments of inertia of these fragments.

In the strongly damped description of $^{235}\mathrm{U}(n_{\rm th},f)$ by Albertsson et al.~\cite{albertsson2020}, the energy available at scission is shared between the deformed prefragments according to their microscopic level densities. The distortion energy is released during shape relaxation after scission. FREYA~\cite{randrup&vogt2021} uses the statistical energy available at scission to populate the transverse modes, with moments of inertia that depend on the assumed shapes. Our temperature comparison asks how much the calculated spins would change if the transverse modes equilibrated with the intrinsic degrees of freedom at the adopted scission configurations. The systematic overestimate of light-fragment spins at $T=1$~MeV disfavors equilibration at that temperature for the adopted inertias and frequencies. A larger intrinsic energy can still coexist with weak thermal population if equilibration is incomplete.

Overall, the comparison supports the zero-point approximation, including for fission of $^{232}\mathrm{Th}$ and $^{238}\mathrm{U}$ induced by neutrons with energies near $2$~MeV. A quantitative estimate of the temperature requires accounting for the changes in fragment angular momentum during neutron and $\gamma$-ray emission.

\FloatBarrier
\section{CONCLUSIONS}\label{sec:conclusion}
	
We calculated spin distributions for primary fragments in low-energy neutron-induced fission of $^{232}\mathrm{Th}$ and $^{238}\mathrm{U}$ and spontaneous fission of $^{252}\mathrm{Cf}$. The mean spins for all three systems reproduce the observed angular-momentum scale and the characteristic sawtooth mass dependence. We also predict spin distributions for the light fragments of $^{238}\mathrm{U}(n,f)$, for which corresponding experimental distributions are currently unavailable.

In the model, fragment spins originate in zero-point bending and wriggling oscillations of the prefragments in the connected configuration near the scission point. Both contributions conserve total angular momentum. The moments of inertia are estimated with a hydrodynamic model from nonequilibrium deformations inferred from prompt-neutron multiplicities. This procedure fixes the inertias independently of the measured spins. The frequencies available for $^{252}\mathrm{Cf}(\mathrm{sf})$ were transferred to the induced reactions. The model uses these frequencies in lieu of calculating individual stiffnesses for each mass split.

The agreement suggests that nonequilibrium prefragment deformations and the associated moments of inertia can account for the overall mass dependence of fragment spin. The lower inertias near closed shells and the higher inertias of more deformed fragments lead to the observed sawtooth pattern. Zero-point bending and wriggling set the scale of angular momentum without appreciable thermal population of these modes. The comparison shows that the zero-point transverse oscillations of the scission configuration can explain the principal features of the observed spin distributions.

Remaining discrepancies may arise from uncertainties in the inferred nonequilibrium deformations, limitations of the hydrodynamic inertias, or changes in spin during prompt neutron and $\gamma$-ray emission. Further work should calculate the bending and wriggling stiffnesses $\kappa_t$ for individual mass splits and fissioning systems and estimate the moments of inertia within a consistent description of the scission geometry. Following the de-excitation of the primary fragments would connect the spins at the scission point to the observed $\gamma$ cascades and provide a more direct test of the model.
	
\section*{Acknowledgements}

This work was supported by the Russian Science Foundation (Grant No.~\mbox{25-22-00697}). We thank Prof. S. G. Kadmensky for discussions of the cold-fission picture and Dr L. V. Titova for helpful comments and constructive contributions.

\begingroup\small\raggedbottom\setlength{\bibsep}{1.25pt}
\bibliographystyle{cpc-unsrt}
\bibliography{bibliography}
\endgroup

\clearpage
\newcommand{\modeshare}[2]{\ensuremath{#1_{\makebox[.65em][l]{\scriptsize #2}}}}
\setcounter{table}{0}
\begin{table*}[p]
\caption{\label{tab:t1}Mean spins from Eq.~\eqref{eq:12} and the corresponding primary-fragment parameters for $^{238}\mathrm{U}(n,f)$.}
		\centering\normalsize\renewcommand{\arraystretch}{1.3}\setlength{\tabcolsep}{3pt}
			\begin{tabular*}{\textwidth}{@{\extracolsep{\fill}}ccccccccccccc}
$^{A_1}X_1$ & $^{A_2}X_2$ & $\beta_{2_1}$ & $\beta_{2_2}$ & $I_b$ & $I_{\rm w}$ & $I_1/I_{\rm rigid}$ & $I_2/I_{\rm rigid}$ & $\hbar\omega_{0_b}$ & $\langle J_1\rangle$ & $\langle J_2\rangle$ & PFF~1 & PFF~2 \\
& & & & $\hbar^2/\text{MeV}$ & $\hbar^2/\text{MeV}$ & & & MeV & $\hbar$ & $\hbar$ & $\%_t$ & $\%_t$ \\
				\hline
${}^{82}\text{Ge}$ & ${}^{157}\text{Nd}$ & 0.372 & 0.723 & 39.79 & 41.77 & 0.10 & 0.55 & 0.7 & 4.69 & 8.56 & \modeshare{99.4}{b} & \modeshare{70.3}{w} \\
${}^{84}\text{Se}$ & ${}^{155}\text{Ce}$ & 0.516 & 0.622 & 38.93 & 41.06 & 0.10 & 0.55 & 0.7 & 4.65 & 8.46 & \modeshare{99.1}{b} & \modeshare{70.0}{w} \\
${}^{86}\text{Se}$ & ${}^{153}\text{Ce}$ & 0.527 & 0.702 & 43.23 & 49.32 & 0.28 & 0.64 & 0.7 & 5.02 & 8.63 & \modeshare{95.2}{b} & \modeshare{68.2}{w} \\
${}^{88}\text{Se}$ & ${}^{151}\text{Ce}$ & 0.536 & 0.690 & 41.30 & 50.48 & 0.42 & 0.62 & 0.7 & 5.09 & 8.13 & \modeshare{88.8}{b} & \modeshare{66.0}{w} \\
${}^{88}\text{Kr}$ & ${}^{151}\text{Ba}$ & 0.533 & 0.663 & 40.53 & 45.68 & 0.24 & 0.60 & 0.7 & 4.80 & 8.33 & \modeshare{95.3}{b} & \modeshare{68.0}{w} \\
${}^{90}\text{Kr}$ & ${}^{149}\text{Ba}$ & 0.561 & 0.660 & 27.49 & 38.39 & 0.42 & 0.40 & 0.7 & 4.46 & 6.31 & \modeshare{75.9}{b} & \modeshare{63.0}{w} \\
${}^{92}\text{Kr}$ & ${}^{147}\text{Ba}$ & 0.602 & 0.650 & 26.30 & 39.34 & 0.47 & 0.40 & 0.8 & 4.92 & 6.45 & \modeshare{69.3}{b} & \modeshare{61.6}{w} \\
${}^{94}\text{Kr}$ & ${}^{145}\text{Ba}$ & 0.632 & 0.637 & 31.70 & 46.80 & 0.53 & 0.50 & 0.8 & 5.37 & 7.06 & \modeshare{70.2}{b} & \modeshare{61.7}{w} \\
${}^{94}\text{Sr}$ & ${}^{145}\text{Xe}$ & 0.627 & 0.588 & 31.09 & 45.99 & 0.53 & 0.50 & 0.8 & 5.34 & 7.08 & \modeshare{69.6}{b} & \modeshare{61.4}{w} \\
${}^{96}\text{Sr}$ & ${}^{143}\text{Xe}$ & 0.659 & 0.594 & 31.04 & 47.54 & 0.55 & 0.50 & 0.85 & 5.63 & 7.08 & \modeshare{65.8}{b} & \modeshare{60.4}{w} \\
${}^{98}\text{Sr}$ & ${}^{141}\text{Xe}$ & 0.676 & 0.587 & 30.17 & 48.29 & 0.58 & 0.50 & 0.91 & 5.93 & 7.03 & \modeshare{61.2}{b} & \modeshare{59.2}{w} \\
${}^{98}\text{Zr}$ & ${}^{141}\text{Te}$ & 0.688 & 0.563 & 29.82 & 48.5 & 0.60 & 0.50 & 0.9 & 5.18 & 7.01 & \modeshare{59.4}{b} & \modeshare{58.6}{w} \\
${}^{100}\text{Zr}$ & ${}^{139}\text{Te}$ & 0.686 & 0.555 & 20.68 & 43.68 & 0.68 & 0.35 & 0.9 & 6.19 & 5.32 & \modeshare{61.0}{w} & \modeshare{52.3}{w} \\
${}^{102}\text{Zr}$ & ${}^{137}\text{Te}$ & 0.717 & 0.546 & 17.03 & 44.34 & 0.77 & 0.30 & 0.9 & 6.67 & 4.64 & \modeshare{72.1}{w} & \modeshare{52.9}{b} \\
${}^{104}\text{Zr}$ & ${}^{135}\text{Te}$ & 0.754 & 0.530 & 16.53 & 45.57 & 0.79 & 0.30 & 0.9 & 6.86 & 4.53 & \modeshare{74.7}{w} & \modeshare{55.0}{b} \\
${}^{102}\text{Mo}$ & ${}^{137}\text{Sn}$ & 0.712 & 0.551 & 16.98 & 35.85 & 0.65 & 0.30 & 0.9 & 6.11 & 4.79 & \modeshare{64.7}{w} & \modeshare{52.9}{b} \\
${}^{104}\text{Mo}$ & ${}^{135}\text{Sn}$ & 0.776 & 0.510 & 16.59 & 41.75 & 0.68 & 0.30 & 0.9 & 6.33 & 4.64 & \modeshare{71.1}{w} & \modeshare{53.1}{b} \\
${}^{130}\text{Sn}$ & ${}^{109}\text{Mo}$ & 0.452 & 0.758 & 12.02 & 35.17 & 0.20 & 0.60 & 1 & 3.66 & 6.29 & \modeshare{63.8}{b} & \modeshare{79.2}{w} \\
${}^{132}\text{Sn}$ & ${}^{107}\text{Mo}$ & 0.470 & 0.754 & 12.26 & 34.59 & 0.20 & 0.60 & 1 & 3.94 & 6.56 & \modeshare{62.4}{b} & \modeshare{77.9}{w} \\
${}^{134}\text{Sn}$ & ${}^{105}\text{Mo}$ & 0.503 & 0.791 & 12.03 & 34.55 & 0.20 & 0.60 & 1 & 4.03 & 6.50 & \modeshare{60.2}{b} & \modeshare{77.7}{w} \\
${}^{132}\text{Te}$ & ${}^{107}\text{Zr}$ & 0.487 & 0.741 & 12.38 & 38.53 & 0.20 & 0.70 & 1 & 3.86 & 7.09 & \modeshare{64.8}{b} & \modeshare{80.8}{w} \\
${}^{134}\text{Te}$ & ${}^{105}\text{Zr}$ & 0.522 & 0.757 & 12.74 & 34.44 & 0.20 & 0.60 & 1 & 4.03 & 6.50 & \modeshare{60.8}{b} & \modeshare{76.4}{w} \\
${}^{136}\text{Te}$ & ${}^{103}\text{Zr}$ & 0.536 & 0.733 & 19.25 & 42.21 & 0.30 & 0.70 & 0.9 & 4.84 & 6.51 & \modeshare{55.5}{b} & \modeshare{67.7}{w} \\
${}^{138}\text{Te}$ & ${}^{101}\text{Zr}$ & 0.543 & 0.702 & 19.69 & 38.44 & 0.30 & 0.60 & 0.9 & 5.11 & 5.99 & \modeshare{51.7}{b} & \modeshare{60.7}{w} \\
${}^{138}\text{Xe}$ & ${}^{101}\text{Sr}$ & 0.565 & 0.679 & 19.60 & 38.35 & 0.30 & 0.60 & 0.9 & 5.10 & 5.97 & \modeshare{51.0}{b} & \modeshare{60.3}{w} \\
${}^{140}\text{Xe}$ & ${}^{99}\text{Sr}$ & 0.582 & 0.672 & 32.14 & 46.91 & 0.50 & 0.55 & 0.9 & 7.22 & 6.00 & \modeshare{57.4}{w} & \modeshare{64.1}{b} \\
${}^{142}\text{Xe}$ & ${}^{97}\text{Sr}$ & 0.593 & 0.671 & 32.82 & 45.82 & 0.50 & 0.50 & 0.85 & 7.26 & 5.63 & \modeshare{59.2}{w} & \modeshare{69.5}{b} \\
${}^{142}\text{Ba}$ & ${}^{97}\text{Kr}$ & 0.616 & 0.649 & 32.67 & 44.63 & 0.50 & 0.50 & 0.85 & 7.25 & 5.63 & \modeshare{59.1}{w} & \modeshare{70.5}{b} \\
${}^{144}\text{Ba}$ & ${}^{95}\text{Kr}$ & 0.620 & 0.639 & 33.19 & 46.95 & 0.50 & 0.50 & 0.8 & 7.18 & 5.38 & \modeshare{61.0}{w} & \modeshare{71.5}{b} \\
${}^{146}\text{Ba}$ & ${}^{93}\text{Kr}$ & 0.620 & 0.592 & 32.80 & 40.66 & 0.50 & 0.50 & 0.8 & 7.30 & 5.29 & \modeshare{59.1}{w} & \modeshare{76.2}{b} \\
${}^{148}\text{Ba}$ & ${}^{91}\text{Kr}$ & 0.659 & 0.580 & 33.02 & 43.67 & 0.50 & 0.40 & 0.7 & 7.14 & 4.67 & \modeshare{64.7}{w} & \modeshare{82.3}{b} \\
${}^{148}\text{Ce}$ & ${}^{91}\text{Se}$ & 0.680 & 0.569 & 33.19 & 43.79 & 0.50 & 0.40 & 0.7 & 7.12 & 4.70 & \modeshare{64.9}{w} & \modeshare{82.7}{b} \\
${}^{150}\text{Ce}$ & ${}^{89}\text{Se}$ & 0.684 & 0.552 & 41.90 & 51.96 & 0.64 & 0.40 & 0.7 & 8.26 & 5.07 & \modeshare{66.6}{w} & \modeshare{89.2}{b} \\
			\end{tabular*}
	\end{table*}

\begin{table*}[p]
\caption{\label{tab:t2}Mean spins from Eq.~\eqref{eq:12} and the corresponding primary-fragment parameters for $^{232}\mathrm{Th}(n,f)$.}
		\centering\normalsize\renewcommand{\arraystretch}{1.3}\setlength{\tabcolsep}{3pt}
			\begin{tabular*}{\textwidth}{@{\extracolsep{\fill}}ccccccccccccc}
$^{A_1}X_1$ & $^{A_2}X_2$ & $\beta_{2_1}$ & $\beta_{2_2}$ & $I_b$ & $I_{\rm w}$ & $I_1/I_{\rm rigid}$ & $I_2/I_{\rm rigid}$ & $\hbar\omega_{0_b}$ & $\langle J_1\rangle$ & $\langle J_2\rangle$ & PFF~1 & PFF~2 \\
& & & & $\hbar^2/\text{MeV}$ & $\hbar^2/\text{MeV}$ & & & MeV & $\hbar$ & $\hbar$ & $\%_t$ & $\%_t$ \\
				\hline
${}^{82}\text{Ge}$ & ${}^{151}\text{Ce}$ & 0.392 & 0.654 & 36.75 & 39.83 & 0.10 & 0.55 & 0.70 & 4.52 & 8.23 & \modeshare{99.2}{b} & \modeshare{70.8}{w} \\
${}^{84}\text{Ge}$ & ${}^{149}\text{Ce}$ & 0.418 & 0.618 & 36.14 & 38.34 & 0.10 & 0.55 & 0.70 & 4.46 & 8.08 & \modeshare{99.1}{b} & \modeshare{70.1}{w} \\
${}^{84}\text{Se}$ & ${}^{149}\text{Ba}$ & 0.415 & 0.599 & 36.57 & 38.72 & 0.10 & 0.55 & 0.70 & 4.47 & 8.12 & \modeshare{99.0}{b} & \modeshare{70.0}{w} \\
${}^{86}\text{Se}$ & ${}^{147}\text{Ba}$ & 0.437 & 0.571 & 32.17 & 38.64 & 0.28 & 0.50 & 0.70 & 4.39 & 7.24 & \modeshare{91.4}{b} & \modeshare{67.0}{w} \\
${}^{88}\text{Se}$ & ${}^{145}\text{Ba}$ & 0.482 & 0.716 & 34.26 & 34.30 & 0.42 & 0.55 & 0.70 & 4.74 & 7.24 & \modeshare{88.4}{b} & \modeshare{59.8}{w} \\
${}^{88}\text{Kr}$ & ${}^{145}\text{Xe}$ & 0.474 & 0.552 & 34.26 & 39.59 & 0.24 & 0.55 & 0.70 & 4.45 & 7.59 & \modeshare{93.5}{b} & \modeshare{67.4}{w} \\
${}^{90}\text{Kr}$ & ${}^{143}\text{Xe}$ & 0.496 & 0.526 & 30.67 & 41.32 & 0.42 & 0.50 & 0.70 & 4.60 & 6.74 & \modeshare{79.5}{b} & \modeshare{63.8}{w} \\
${}^{92}\text{Kr}$ & ${}^{141}\text{Xe}$ & 0.52 & 0.525 & 29.51 & 42.51 & 0.47 & 0.50 & 0.80 & 5.05 & 6.92 & \modeshare{73.6}{b} & \modeshare{62.8}{w} \\
${}^{94}\text{Kr}$ & ${}^{139}\text{Xe}$ & 0.55 & 0.521 & 23.38 & 39.95 & 0.56 & 0.40 & 0.80 & 5.17 & 5.71 & \modeshare{56.4}{b} & \modeshare{58.5}{w} \\
${}^{92}\text{Sr}$ & ${}^{141}\text{Te}$ & 0.518 & 0.487 & 29.40 & 36.69 & 0.30 & 0.50 & 0.80 & 4.59 & 7.31 & \modeshare{85.8}{b} & \modeshare{64.9}{w} \\
${}^{94}\text{Sr}$ & ${}^{139}\text{Te}$ & 0.539 & 0.502 & 17.82 & 33.76 & 0.53 & 0.30 & 0.80 & 4.91 & 4.84 & \modeshare{52.6}{w} & \modeshare{55.8}{w} \\
${}^{96}\text{Sr}$ & ${}^{137}\text{Te}$ & 0.649 & 0.436 & 17 & 37.45 & 0.64 & 0.30 & 0.85 & 5.64 & 4.73 & \modeshare{63.8}{w} & \modeshare{50.9}{w} \\
${}^{98}\text{Sr}$ & ${}^{135}\text{Te}$ & 0.719 & 0.454 & 16.50 & 35.91 & 0.58 & 0.30 & 0.90 & 5.65 & 4.84 & \modeshare{63.3}{w} & \modeshare{51.0}{w} \\
${}^{98}\text{Zr}$ & ${}^{135}\text{Sn}$ & 0.731 & 0.437 & 16.56 & 36.51 & 0.60 & 0.30 & 0.90 & 5.72 & 4.82 & \modeshare{64.5}{w} & \modeshare{50.0}{} \\
${}^{100}\text{Zr}$ & ${}^{133}\text{Sn}$ & 0.735 & 0.443 & 10.94 & 35.53 & 0.68 & 0.20 & 0.90 & 6.22 & 3.58 & \modeshare{80.3}{w} & \modeshare{59.2}{b} \\
${}^{130}\text{Sn}$ & ${}^{103}\text{Zr}$ & 0.411 & 0.774 & 9.16 & 30.89 & 0.15 & 0.60 & 1 & 3.28 & 6.43 & \modeshare{67.9}{b} & \modeshare{83.1}{w} \\
${}^{132}\text{Sn}$ & ${}^{101}\text{Zr}$ & 0.358 & 0.766 & 12.41 & 32.70 & 0.20 & 0.60 & 1 & 4.00 & 6.27 & \modeshare{62.7}{b} & \modeshare{76.3}{w} \\
${}^{132}\text{Te}$ & ${}^{101}\text{Sr}$ & 0.371 & 0.740 & 12.35 & 32 & 0.20 & 0.60 & 1 & 4.00 & 6.25 & \modeshare{62.5}{b} & \modeshare{75.8}{w} \\
${}^{134}\text{Te}$ & ${}^{99}\text{Sr}$ & 0.420 & 0.727 & 12.59 & 32 & 0.30 & 0.60 & 0.90 & 4.97 & 5.87 & \modeshare{52.5}{w} & \modeshare{68.3}{w} \\
${}^{136}\text{Te}$ & ${}^{97}\text{Sr}$ & 0.430 & 0.68 & 19.01 & 36.43 & 0.30 & 0.60 & 0.90 & 5.07 & 5.77 & \modeshare{52.1}{b} & \modeshare{60.2}{w} \\
${}^{138}\text{Te}$ & ${}^{95}\text{Sr}$ & 0.455 & 0.627 & 19.10 & 35.79 & 0.30 & 0.60 & 0.95 & 5.01 & 5.42 & \modeshare{50.0}{} & \modeshare{57.5}{w} \\
${}^{138}\text{Xe}$ & ${}^{95}\text{Kr}$ & 0.473 & 0.623 & 19.26 & 35.99 & 0.30 & 0.60 & 0.85 & 5.02 & 5.48 & \modeshare{50.2}{w} & \modeshare{57.2}{w} \\
${}^{140}\text{Xe}$ & ${}^{93}\text{Kr}$ & 0.516 & 0.561 & 30.69 & 42.97 & 0.50 & 0.50 & 0.85 & 7.15 & 5.38 & \modeshare{60.3}{w} & \modeshare{71.1}{b} \\
${}^{142}\text{Xe}$ & ${}^{91}\text{Kr}$ & 0.517 & 0.514 & 30.93 & 43.27 & 0.50 & 0.50 & 0.80 & 7.09 & 5.11 & \modeshare{61.6}{w} & \modeshare{73.3}{b} \\
${}^{142}\text{Ba}$ & ${}^{91}\text{Se}$ & 0.535 & 0.507 & 32.09 & 43.24 & 0.50 & 0.50 & 0.80 & 7.08 & 5.15 & \modeshare{60.9}{w} & \modeshare{74.4}{b} \\
${}^{144}\text{Ba}$ & ${}^{89}\text{Se}$ & 0.558 & 0.491 & 32.46 & 41.13 & 0.50 & 0.40 & 0.70 & 6.91 & 4.58 & \modeshare{63.6}{w} & \modeshare{82.6}{b} \\
${}^{146}\text{Ba}$ & ${}^{87}\text{Se}$ & 0.565 & 0.459 & 31.61 & 41.30 & 0.50 & 0.40 & 0.70 & 7.02 & 4.54 & \modeshare{65.2}{w} & \modeshare{83.8}{b} \\
${}^{148}\text{Ce}$ & ${}^{85}\text{Ge}$ & 0.594 & 0.435 & 35.61 & 44.73 & 0.55 & 0.40 & 0.70 & 7.53 & 4.69 & \modeshare{66.1}{w} & \modeshare{87.7}{b} \\
${}^{150}\text{Ce}$ & ${}^{83}\text{Ge}$ & 0.640 & 0.418 & 40.70 & 50.33 & 0.64 & 0.40 & 0.70 & 8.26 & 4.87 & \modeshare{67.9}{w} & \modeshare{91.5}{b} \\
			\end{tabular*}
	\end{table*}

\begin{table*}[p]
\caption{\label{tab:t3}Mean spins from Eq.~\eqref{eq:12} and the corresponding primary-fragment parameters for $^{252}\mathrm{Cf}(\mathrm{sf})$.}
		\centering\normalsize\renewcommand{\arraystretch}{1.3}\setlength{\tabcolsep}{3pt}
			\begin{tabular*}{\textwidth}{@{\extracolsep{\fill}}ccccccccccccc}
$^{A_1}X_1$ & $^{A_2}X_2$ & $\beta_{2_1}$ & $\beta_{2_2}$ & $I_b$ & $I_{\rm w}$ & $I_1/I_{\rm rigid}$ & $I_2/I_{\rm rigid}$ & $\hbar\omega_{0_b}$ & $\langle J_1\rangle$ & $\langle J_2\rangle$ & PFF~1 & PFF~2 \\
& & & & $\hbar^2/\text{MeV}$ & $\hbar^2/\text{MeV}$ & & & MeV & $\hbar$ & $\hbar$ & $\%_t$ & $\%_t$ \\
				\hline
${}^{96}\text{Sr}$ & ${}^{156}\text{Nd}$ & 0.588 & 0.892 & 45.70 & 62 & 0.58 & 0.63 & 0.80 & 6.12 & 8.88 & \modeshare{80.7}{b} & \modeshare{65.9}{w} \\
${}^{98}\text{Sr}$ & ${}^{154}\text{Nd}$ & 0.639 & 0.867 & 45.20 & 62.20 & 0.58 & 0.64 & 0.80 & 6.15 & 8.79 & \modeshare{78.6}{b} & \modeshare{65.2}{w} \\
${}^{98}\text{Zr}$ & ${}^{154}\text{Ce}$ & 0.645 & 0.864 & 42.30 & 59 & 0.55 & 0.60 & 0.90 & 6.35 & 9.06 & \modeshare{78.0}{b} & \modeshare{65.3}{w} \\
${}^{100}\text{Zr}$ & ${}^{152}\text{Ce}$ & 0.697 & 0.813 & 39.20 & 60.10 & 0.67 & 0.56 & 0.90 & 6.63 & 8.35 & \modeshare{66.0}{b} & \modeshare{62.0}{w} \\
${}^{102}\text{Zr}$ & ${}^{150}\text{Ce}$ & 0.740 & 0.759 & 37 & 59.50 & 0.69 & 0.54 & 0.90 & 6.71 & 7.96 & \modeshare{60.0}{b} & \modeshare{60.1}{w} \\
${}^{104}\text{Zr}$ & ${}^{148}\text{Ce}$ & 0.740 & 0.759 & 33.30 & 56.70 & 0.68 & 0.50 & 0.90 & 6.67 & 7.44 & \modeshare{55.2}{b} & \modeshare{59.2}{w} \\
${}^{102}\text{Mo}$ & ${}^{150}\text{Ba}$ & 0.763 & 0.715 & 34.40 & 51.87 & 0.53 & 0.50 & 0.90 & 6.09 & 7.91 & \modeshare{65.9}{b} & \modeshare{61.4}{w} \\
${}^{104}\text{Mo}$ & ${}^{148}\text{Ba}$ & 0.775 & 0.668 & 33.30 & 56.10 & 0.66 & 0.50 & 0.90 & 6.6 & 7.47 & \modeshare{54.6}{b} & \modeshare{58.2}{w} \\
${}^{106}\text{Mo}$ & ${}^{146}\text{Ba}$ & 0.810 & 0.615 & 32.29 & 53.66 & 0.60 & 0.50 & 0.90 & 6.41 & 7.4 & \modeshare{54.8}{b} & \modeshare{57.8}{w} \\
${}^{108}\text{Mo}$ & ${}^{144}\text{Ba}$ & 0.845 & 0.604 & 31.50 & 49.95 & 0.50 & 0.50 & 0.90 & 6.07 & 7.45 & \modeshare{58.9}{b} & \modeshare{58.9}{w} \\
${}^{108}\text{Ru}$ & ${}^{144}\text{Xe}$ & 0.900 & 0.567 & 31.49 & 49.93 & 0.51 & 0.50 & 1 & 6.4 & 7.85 & \modeshare{57.1}{b} & \modeshare{57.5}{w} \\
${}^{110}\text{Ru}$ & ${}^{142}\text{Xe}$ & 0.938 & 0.559 & 24.84 & 47.41 & 0.70 & 0.40 & 1 & 6.65 & 6.57 & \modeshare{62.5}{w} & \modeshare{53.3}{b} \\
${}^{112}\text{Ru}$ & ${}^{140}\text{Xe}$ & 0.444 & 0.997 & 23.94 & 47.23 & 0.70 & 0.40 & 1.20 & 7.35 & 7.00 & \modeshare{54.0}{w} & \modeshare{58.2}{w} \\
${}^{112}\text{Pd}$ & ${}^{140}\text{Te}$ & 0.498 & 0.976 & 23.94 & 47.27 & 0.70 & 0.40 & 1.20 & 7.36 & 7.00 & \modeshare{55.0}{w} & \modeshare{57.4}{w} \\
${}^{114}\text{Pd}$ & ${}^{138}\text{Te}$ & 0.549 & 0.914 & 17.25 & 41.98 & 0.70 & 0.30 & 1.20 & 7.36 & 5.62 & \modeshare{68.5}{w} & \modeshare{52.0}{w} \\
${}^{116}\text{Pd}$ & ${}^{136}\text{Te}$ & 0.559 & 0.938 & 16.85 & 42.32 & 0.70 & 0.30 & 1.20 & 7.52 & 5.51 & \modeshare{70.0}{w} & \modeshare{51.5}{w} \\
${}^{130}\text{Sn}$ & ${}^{122}\text{Cd}$ & 0.591 & 0.865 & 10.26 & 27.89 & 0.20 & 0.40 & 1.10 & 4.15 & 5.84 & \modeshare{57.0}{b} & \modeshare{76.2}{w} \\
${}^{132}\text{Sn}$ & ${}^{120}\text{Cd}$ & 0.604 & 0.845 & 11.36 & 28.97 & 0.20 & 0.40 & 1.10 & 4.26 & 6.05 & \modeshare{56.3}{b} & \modeshare{74.2}{w} \\
${}^{134}\text{Sn}$ & ${}^{118}\text{Cd}$ & 0.615 & 0.817 & 10.45 & 30.73 & 0.20 & 0.50 & 1.10 & 4.15 & 6.22 & \modeshare{57.8}{b} & \modeshare{78.5}{w} \\
${}^{136}\text{Te}$ & ${}^{116}\text{Pd}$ & 0.668 & 0.775 & 15.80 & 42.32 & 0.30 & 0.70 & 1.20 & 5.42 & 7.45 & \modeshare{54.5}{b} & \modeshare{74.8}{w} \\
${}^{138}\text{Xe}$ & ${}^{114}\text{Ru}$ & 0.681 & 0.763 & 16.17 & 41.98 & 0.30 & 0.70 & 1.20 & 5.53 & 7.34 & \modeshare{53.2}{b} & \modeshare{73.4}{w} \\
${}^{140}\text{Xe}$ & ${}^{112}\text{Ru}$ & 0.759 & 0.740 & 21.79 & 47.23 & 0.40 & 0.70 & 1.20 & 6.85 & 7.22 & \modeshare{53.9}{w} & \modeshare{62.6}{w} \\
${}^{142}\text{Xe}$ & ${}^{110}\text{Ru}$ & 0.794 & 0.729 & 22.32 & 47.41 & 0.40 & 0.70 & 1 & 6.42 & 6.5 & \modeshare{55.4}{w} & \modeshare{60.4}{w} \\
${}^{142}\text{Ba}$ & ${}^{110}\text{Mo}$ & 0.857 & 0.673 & 25.23 & 50.25 & 0.40 & 0.70 & 0.90 & 6.32 & 6.6 & \modeshare{55.4}{w} & \modeshare{57.4}{w} \\
${}^{144}\text{Ba}$ & ${}^{108}\text{Mo}$ & 0.864 & 0.645 & 33.55 & 49.95 & 0.50 & 0.50 & 0.90 & 7.54 & 6.19 & \modeshare{61.6}{w} & \modeshare{67.0}{b} \\
${}^{146}\text{Ba}$ & ${}^{106}\text{Mo}$ & 0.864 & 0.645 & 34.54 & 53.66 & 0.50 & 0.60 & 0.90 & 7.51 & 6.34 & \modeshare{60.3}{w} & \modeshare{62.3}{b} \\
${}^{148}\text{Ce}$ & ${}^{104}\text{Zr}$ & 0.794 & 0.729 & 35.58 & 57.39 & 0.50 & 0.70 & 0.90 & 7.52 & 6.85 & \modeshare{57.7}{w} & \modeshare{56.4}{b} \\
${}^{150}\text{Ce}$ & ${}^{102}\text{Zr}$ & 0.857 & 0.673 & 45.36 & 65.24 & 0.64 & 0.70 & 0.90 & 8.85 & 7.01 & \modeshare{61.7}{w} & \modeshare{69.3}{b} \\
${}^{152}\text{Nd}$ & ${}^{100}\text{Sr}$ & 0.864 & 0.645 & 43.22 & 60.21 & 0.60 & 0.60 & 0.90 & 8.86 & 6.59 & \modeshare{63.4}{w} & \modeshare{74.4}{b} \\
${}^{154}\text{Nd}$ & ${}^{98}\text{Sr}$ & 0.864 & 0.645 & 47.90 & 63.32 & 0.66 & 0.58 & 0.80 & 8.99 & 6.28 & \modeshare{64.7}{w} & \modeshare{79.9}{b} \\
			\end{tabular*}
	\end{table*}
\clearpage

\end{document}